\documentclass[twocolumn]{svjour3}          
\smartqed  
\usepackage{graphicx}
\usepackage{amsmath}
\usepackage{color}
\usepackage{epstopdf}
\usepackage[square,sort,comma,numbers]{natbib}
\usepackage[affil-it]{authblk}
 \usepackage{mathptmx}      
\newcommand{\textunderscript}[1]{$_{\text{#1}}$}

 \journalname{Engineering Fluid Dynamics Group Report}
\begin{document}

\title{Schlieren study of a sonic jet injected into a supersonic cross flow using high-current pulsed LEDs 
}


\author{Ella Giskes         \and
        Ruben A. Verschoof \and
        Frans B. Segerink \and
        Cornelis H. Venner 
}


\institute{E. Giskes \and C.H. Venner \at
              Engineering Fluid Dynamics group, Faculty of Mechanical Engineering, University of Twente, P.O. Box 217, 7500AE, Enschede, the Netherlands \\
              \email{e.giskes@utwente.nl}  \\         
              \email{c.h.venner@utwente.nl}    \\       
           \and
           R.A. Verschoof \at
              Physics of Fluids group, MESA+ institute, J.M. Burgers Centre for Fluid Dynamics, University of Twente, P.O. Box 217, 7500AE Enschede, the Netherlands \\
                         \and
           F.B. Segerink \at
              Optical Sciences group, MESA+ institute, University of Twente, P.O. Box 217, 7500AE, Enschede, the Netherlands \\
}

\date{Received: date / Accepted: date}

\maketitle

\begin{abstract}
Benefiting from the development of increasingly advanced high speed cameras, flow visualization and analysis nowadays yield detailed data of the flow field in many applications. Notwithstanding this progress, for high speed and supersonic flows it is still not trivial to capture high quality images. In this study we present a Schlieren setup that uses pulsed LEDs with high currents (up to 18 Ampere) to increase the optical output to sufficient levels. The bright and short pulses, down to 130 nanoseconds, allow detailed and sharp imaging of the flow with a high spatial resolution adequate for supersonic flow. The pulse circuit and pulse width determination are explained in detail.
As a  test case we studied the near field of a 2 mm diameter sonic jet injected transversely into a supersonic cross flow. This is a model flow for fuel injection in scramjet engines, which is not yet fully understood. Owing to the high resolution and accuracy of the images produced by the newly developed system we prove the existence of a large (density) gradient wave traveling in the windward subsonic region between the Mach barrel and the bowshock, which hitherto was observed only in some numerical studies but not yet shown in experiments. Furthermore, we demonstrate with this Schlieren setup that time-correlated images can be obtained, with an interframe time of 2 microseconds, so that also flow unsteadiness can be studied such as the movement of shock waves and trajectories of vortices. The obtained results of the jet penetration height are presented as a power law correlation.
The results of this study show that the designed setup using a low-cost LED and low-cost control system is a high potential option for application in visualization studies of high speed flows.

\keywords{Supersonic flows \and jets \and mixing \and turbulence \and LED \and Schlieren \and Rogowski coil}
\end{abstract}

\section{Introduction}
\label{intro}
In the study of high-speed compressible flows it is still hard to obtain quantitative flow properties by non-intrusive means. Flow visualization remains an important experimental tool (\citet{Mahesh2013}). The unsteady behavior of high speed flows requires high spatial and temporal resolution to resolve and follow key events in the flow field in detail. Particularly, when smaller structures are to be studied, the resolution requirements and therefore also the requirements for pulse duration and illumination brightness, are high. 

The injection of a jet into a supersonic cross flow, used as a model for fuel injection in a scramjet engine, is an example of such a high speed and unsteady flow field (\citet{Ben-Yakar2006,Vanlerberghe2000,Santiago1997,Papamoschou1993}). Efficient fuel injection is necessary for a successful engine design, therefore fundamental understanding of the flow field is essential. The occurring structures in the flow are widely agreed on, however, the mechanism which governs their behaviour is not yet fully known (\citet{Ben-Yakar2006,Portz2006,Gamba2014,Mahesh2013}). The high velocities of the flow field have limited the amount of available experimental data to date. This also affects numerical efforts for this type of flow which require experimental data for code validation, because for all practical numerical studies turbulence modelling is necessary. More high-quality experimental data is therefore required to better understand the flow at a fundamental level (\citet{Bertin2003,Fulton2014,Rockwell2014,Mahesh2013}).

The present study focussed on developing a Schlieren setup using an overdriven LED to generate an easy-to-use and a cost-efficient technique that visualizes and tracks density gradients in a supersonic flow field. For details on the Schlieren technique we refer to the classical textbook by \citet{Settles2001}. LED light outperforms other light sources in a Schlieren setup such as Xenon flashbulbs and lasers, both in terms of image quality and ease of operation (\citet{Willert2010,Settles2001}). The typical  bandwidth of the emitted LED light results in high quality Schlieren images. Due to the nature of Schlieren, which is based on the refraction of light, very small bandwidths, obtained with e.g. pulsed lasers, see \citet{Parziale2015}, result in speckle patterns in the Schlieren image and a large bandwidth, obtained with e.g. Xenon flashbulbs, results in smeared refraction and less sharp images for equal settings in the setup. Also, the typical size and square shape of LED lights simplifies a Schlieren setup, because the use of a condenser lens/pin-hole diaphragm combination is not necessary.  In addition, the fast response time of LEDs to a supplied current increases the flexibility of a setup. Light output of LEDs has been a limitation. However, LEDs have improved so much over the past years that the use of LEDs has become feasible for high-speed Schlieren imaging (\citet{Willert2012,Wilson2015}). In these publications it was shown that the technique can be used to study a 5 mm diameter sonic jet in {\em still air}.  However, the case of a jet in a supersonic cross flow is much more demanding.

We consider  a supersonic cross flow and a jet injection nozzle of only 2 mm diameter. Single shot time-instantaneous images obtained with e.g. flashbulbs do exist for a sonic jet in a supersonic cross flow. However, regarding LED based Schlieren imaging, a sonic jet in a supersonic cross flow requires a much higher spatial and temporal resolution than, to the knowledge of the authors, has been demonstrated so far in the literature where LED was used. The developed setup yields sharp images of the flow structures with a high resolution. In this article we present not only the temporal and spatial resolutions obtained, but also other relevant settings of the Schlieren setup. Therefore, this article provides all information required to rebuild this setup for use in other studies in other facilities.
  \begin{figure}[!htbp]
\centering
	\includegraphics[width=0.50\textwidth]{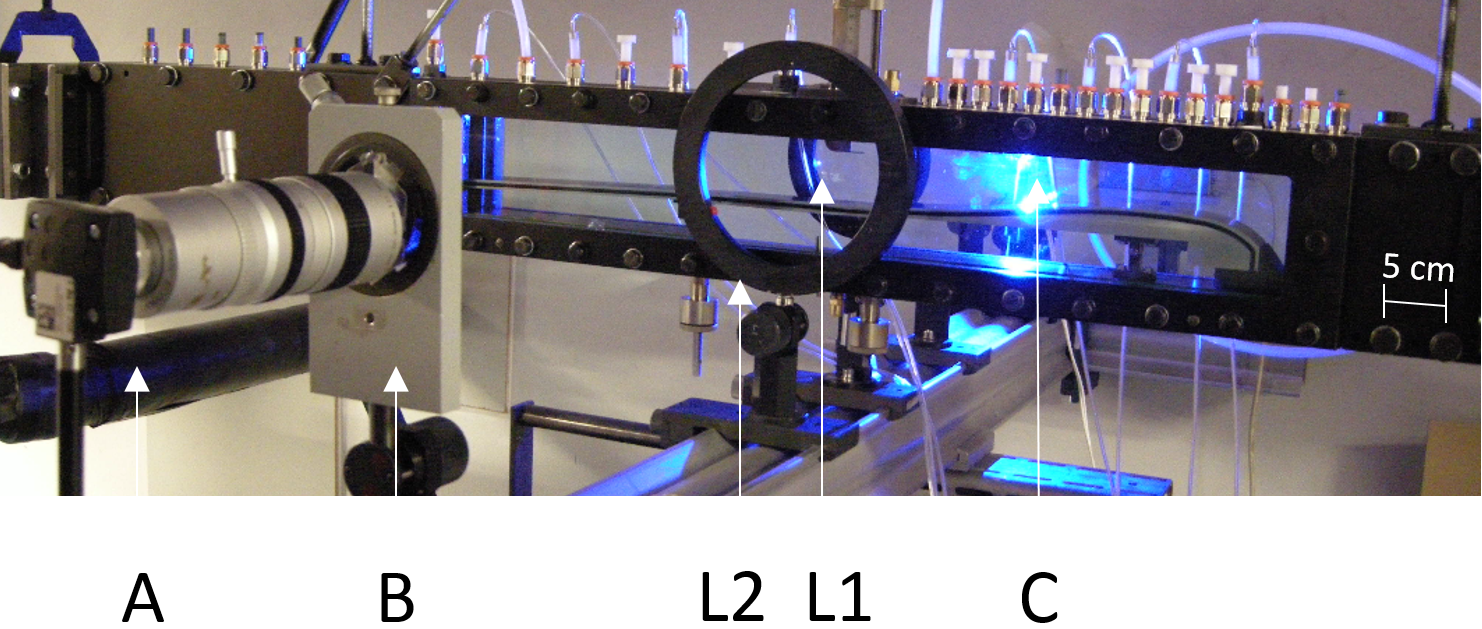}
\caption{Photograph of the wind tunnel and Schlieren setup. The LED (C) emits light towards Lens 1 (L1) after which the light travels in a parallel beam. The light hits the first glass window perpendicularly after which it travels through the test section and leaves the test section to hit Lens 2 (L2). The light is focussed by Lens 2 on a knife edge (B) and the remaining light bundle is captured by a camera (A). The wind tunnel comprises glass windows and a steel flat upper and curved steel lower wall. From the upper wall a jet is injected into the test section. The direction of the wind tunnel flow is from right to left.}
\label{fig:setup}       
\end{figure}
The layout of this article is as follows. i) We first describe the experimental setup: the wind tunnel and the jet, the Schlieren setup and particularly the design and performance of the LED driver. ii) We show and discuss results of our test case, i.e. the injection of a sonic jet in a supersonic cross flow, and present the jet penetration height described by a `jet penetration relation'.


\section{Experimental Setup}
\subsection{Wind tunnel and jet injection}
The experiments have been conducted in the Supersonic Wind Tunnel at the University of Twente, shown in figure \ref{fig:setup}. The wind tunnel is an open-return wind tunnel with an adjustable nozzle. The inlet of the tunnel is at atmospheric conditions, consequently it provides a low enthalpy flow of $h_0=0.3$ MJ/kg for air at $T_0=293$K. At the location of the test section, the free-stream cross flow had a Mach number of M $=1.7\pm0.02$, corresponding to a velocity of $475\pm10$ m/s, a static pressure of $p=21.3\pm2$ kPa, and a static temperature of $T=195\pm10$ K. The boundary layer is a fully-developed turbulent boundary layer. The boundary layer thickness equals $\delta_{99}=3.8$ mm, based at the location where the velocity is $99\%$ of the free stream velocity. The displacement thickness equals $\delta_1=0.84$ mm. The unit Reynolds number in the experiments  was $1.4 \cdot 10^7$   per meter based on the free stream quantities in the test section (M=1.7) (\citet{Rouwenhorst2012}).

The test section comprises a cross section approximately 200~mm downstream of the throat of the wind tunnel nozzle. At this location, the cross-section is 45~mm wide by 60~mm high. The flow properties are practically uniform in this cross section. Optical access to the test section is provided by glass windows in the side walls.

Through a jet-nozzle with an outlet diameter of 2~mm, pressurized air was injected at a fixed stagnation temperature of $T_{0,j}$~=~293~K. The jet nozzle is converging and the ratio of the pressure in the jet reservoir and the free-stream pressure was for all experiments such that the jet was sonic at its outlet resulting in an underexpanded sonic jet. The static pressure of the jet was measured a small distance (65~mm) upstream of the jet orifice in a 10~mm diameter supply plenum by a sensor of a 15 psi range Netscanner 9116 pressure systems box for pressures between 0 and 200 kPa and a GE Druck DPI 104 for higher pressures. This static pressure has been related to the jet reservoir pressure as well as the static jet pressure at the orifice by the relations for isentropic flow.

\subsection{Schlieren system}
This study uses Schlieren imaging to visualize density gradients in the supersonic cross flow. A photograph of the in-line Schlieren setup is included in figure \ref{fig:setup}. The setup comprises a LED, type Luxeon SP-05-L1, as light source, lens L1 with focal length 500~mm and a diameter of 88~mm, lens L2 with focal length 1000~mm and a diameter of 146~mm and a  knife edge. The small, square shaped LED simplified the setup, i.e. the use of a condenser lens and pin-hole diaphragm necessary for a conventional light source was not required. The LED was placed at 0.5 m distance from the first lens. These choices improved the luminous flux through the setup as compared to previous studies (\citet{Willert2012, Wilson2015, Morimoto2016}), increasing significantly the spatial and temporal resolutions. 
Furthermore, theoretically, the magnitude of the detectable density variations is inversely proportional to the focal length of L2. As such one would generally aim for a large focal length. However, we show that with the developed setup and the utilized cutoff percentages, a sharp image of the flow is obtained with great detail without the need of a long focal length.  All relevant flow structures of the flow interaction field are captured. We used two different cameras: a Photron SA7 CCD (1024x1024 pixel) camera in combination with both a Zeiss 100~mm lens and a Zeiss 400~mm lens to obtain single images, and to obtain sets of two time-correlated images a pco pco.edge 5.5 sCMOS (2560x2160 pixel) camera in combination with a Zeiss 100~mm lens. 

\begin{figure}
\centering
\includegraphics[width=0.50\textwidth]{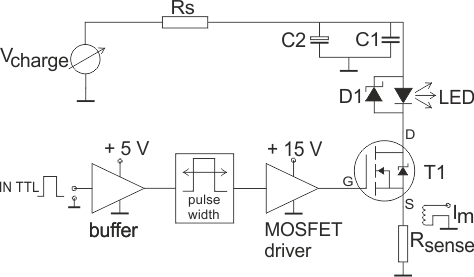}
\caption{Schematic of the deployed in-house built LED driver. Details of all components are shown in table \ref{tab:1}}.
\label{fig:driver}       
\end{figure}
\begin{table}
\caption{LED driver components and parameters}
\label{tab:1}       
\begin{tabular}{ll}
\hline\noalign{\smallskip}
Rs &  20 $\Omega$ \\
C1  &   10$\mu$F SMD, 6 $\times$ parallel\\
C2   &  330 $\mu$F, EEEFT1H331AP, 6 $\times$ parallel \\
D1   &  BYQ28 \\
T1   &  IRFB3206PBF \\
LED &  SP-05-L1, Luxeon \\
R\textunderscript{sense} & two times 0.02 $\Omega$ parallel \\
I\textunderscript{m} & External circuit with Rogowski coil, see text \\
MOSFET driver & EL7104  \\ 
Buffer  & EL7104  \\
V\textunderscript{charge} & LT3086, adjustable voltage regulator  \\
Pulse width circuit   & See text  \\
\noalign{\smallskip}\hline
\end{tabular}
\end{table}

\subsection{LED driver}
{The LED light was controlled by an in-house built driver (figure \ref{fig:driver}, table \ref{tab:1}), which allowed the optical pulse width and its brightness to be adjusted independently. The pulse is TTL triggered and therefore it can be synchronised with the camera.

Driving LEDs with a high current in short pulse operation can be achieved with only a few components, specified in table 1. As shown in figure \ref{fig:driver}, capacitor banks C1 and C2 are charged by an adjustable voltage source V\textunderscript{charge}. The MOSFET T1 acts as a near ideal switch when triggered. Depending on the charge voltage, it results in a fast and high current pulse. The speed of the circuit at given charge voltage is primarily dependent on the applied MOSFET and on the circuit layout. In this case, currents up to 840 A are permissible for the applied MOSFET. Rise and fall times are only a few tens of nanoseconds, so the vast majority of LEDs can be driven to the edge of their specifications and beyond. 
Resistor Rs limits the charge current and the maximum average current.  Capacitor C2 consists of 6 parallel capacitors of 330~$\mu$F, allowing long duration pulses or pulse trains. Paralleling capacitors lowers their series resistance and inductance. C1 is 6 times 10~$\mu$F SMD, parallel, which is placed close to the LED, further minimizing impedance, allowing faster pulses.
Shottky diode D1 prevents reverse currents from damaging the LED.

As the circuit is capable of delivering a current with $di/dt$ of about $10^{10}$~A/s, circuit layout is critical and connections have to be kept as short as possible. In case of a mounted LED, some wiring is unavoidable, so some ringing of the current will always occur.

For high frequency current measurements, the Rogowski coil is a simple and low cost solution. A Rogowski coil picks up magnetic flux variations produced by the pulse current and converts it into a voltage. Argueso et al. (\citet{Argueso2005}) show a simple design, resulting in a true current output for high frequencies, instead of an output proportional to the derivative of the current. This is achieved by terminating the coil with a (very) low impedance, in our case a (virtual) shortcircuit connecting the coil directly to the inverting input of a high bandwidth operational amplifier. A home built setup is shown in figure \ref{fig:coil}. Results are given in \ref{fig:coil2}, showing a high sensitivity of approximately 2V/A with the coil in close proximity of the current wire.

\begin{figure}[!htbp]
\centering
  \includegraphics[width=0.5\textwidth]{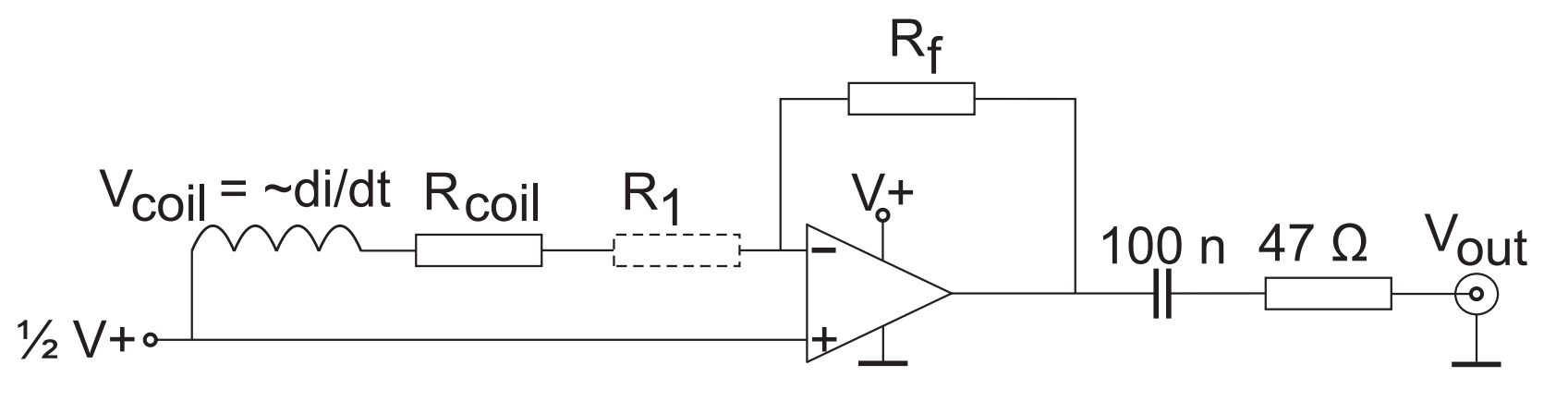}
\caption{Schematic of the developed Rogowski coil circuit, to be used for current monitoring. The circuit is fully independent of the pulser. The coil consists of a miniature surface mounted device in a 0805 housing, measuring 2~mm x 1.2~mm x 1.3~mm (WxLxH). The coil and electronics are built on a separate small circuit board and can be positioned near a current conducting wire, where it will pick up magnetic field variations associated with the current. Due to the low impedance termination of the coil, the circuit acts as a self-integrating Rogowski coil and will output the momentary current.}
\label{fig:coil}       
\end{figure}
\begin{table}[!htbp]
\caption{Components and parameters of the Rogowski coil assembly}
\label{tab:2}       
\begin{tabular}{ll}
\hline\noalign{\smallskip}
Coil & type B82498B (TDK), DC resistance 1.9 $\Omega$,\\
 & inductance 470 nH, self resonant frequency\\
 & 650 MHz, housing 0805, 2 mm x 1.2 mm x 1.3 mm\\
 &  (WxLxH)\\
Op-amp & THS 3091 (Texas Instruments), bandwidth 210 MHz\\
R\textunderscript{coil} & 1.9 $\Omega$ internal DC-resistance of the coil\\
R\textunderscript{1} & 300 $\Omega$, to insert if the derivative of the current is to\\
 & be measured\\
R\textunderscript{f} & 1.2 k$\Omega$ gain setting resistor\\
100 n/47 $\Omega$ & ac-coupling to 50 $\Omega$ output\\
\noalign{\smallskip}\hline
\end{tabular}
\end{table}

For testing a function generator (Hameg HM8030-4) was connected to an amplifier driving a 10 $\Omega$ resistive load (SMD chip, type WSMHP20), connected in close proximity of the output. A square wave of 3.7 MHz was applied, resulting in 360 mA peak current, represented by the black curve in figure \ref{fig:coil2}.

\begin{figure}[!htbp]
\centering
  \includegraphics[width=0.52\textwidth]{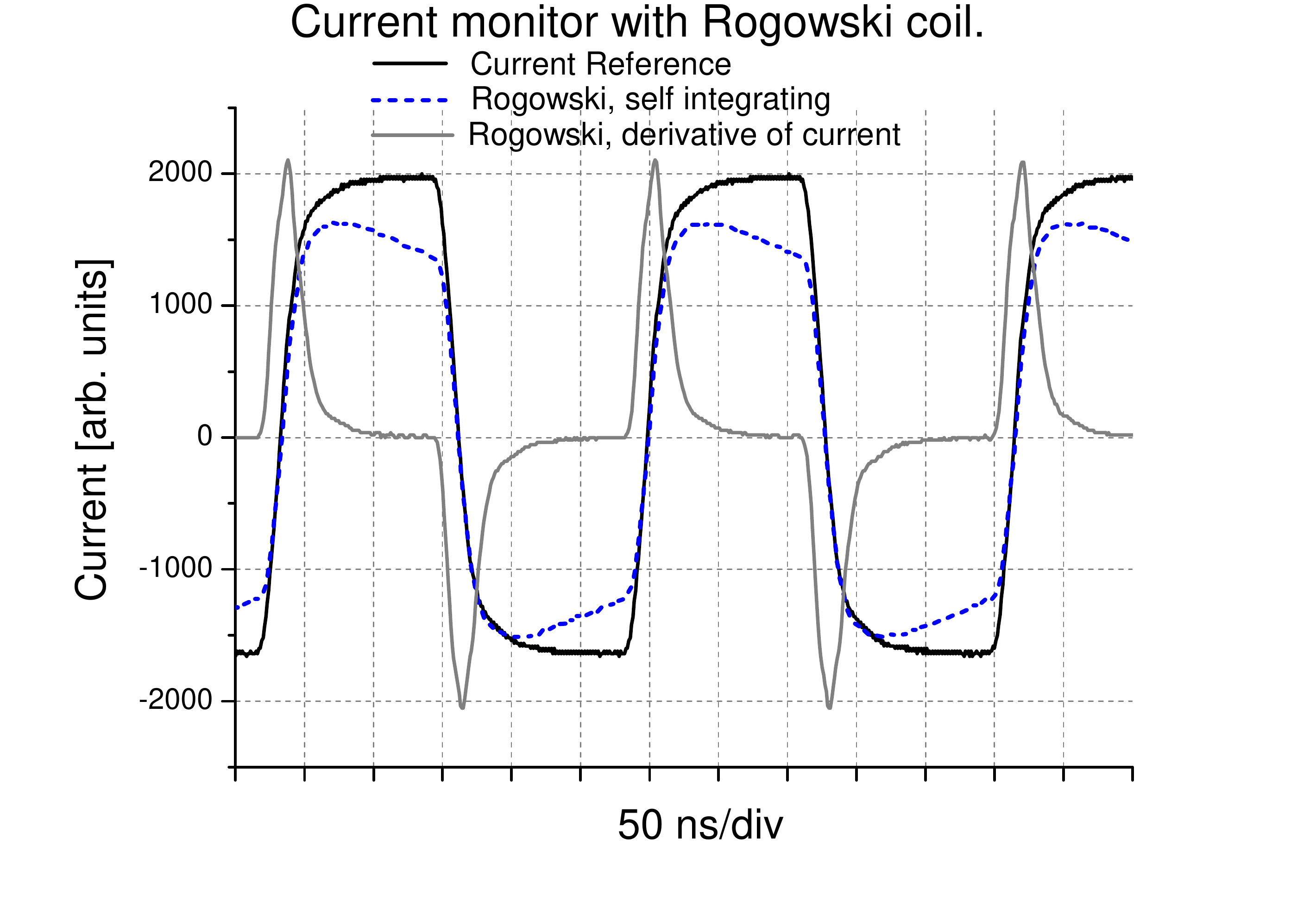}
\caption{Test results for the Rogowski coil. The vertical axis is in arbitrary units to combine several traces. Black trace: voltage over the 10 $\Omega$ load resistor, resulting in a peak current of 360 mA. Blue curve: Output voltage of the Rogowski module, 750 mV peak, sensitivity = 2.0 V/A. The grey curve is obtained when R1 = 300 $\Omega$ is inserted, creating a high impedance load for the coil and thus shifts the Rogowski setup completely to the non-integrating regime. Traces are recorded on a OWON DS8302 oscilloscope, 300 MHz bandwidth.}
\label{fig:coil2}       
\end{figure}

The blue curve represents the basic circuit drawn in \ref{fig:coil} (R1=0 $\Omega$), with a coil of 470 nH. Comparable results (not shown here) were obtained with lower inductance coils and proportionally lower DC resistances, even down to 10 nH. The result clearly shows the self-integrating behavior of the Rogowski module roughly above 1~MHz, where the impedance of the coil ($\omega$L) gets bigger than the coils DC-resistance. For an explanation see Argueso et al. (\citet{Argueso2005}). The grey curve demonstrates the effect of a high impedance termination of the coil. A value of R1 = 300~$\Omega$ equals ωL at 100~MHz, so up to 100~MHz this is to be considered a high impedance. Figure \ref{fig:coil2} (grey curve) clearly shows that the Rogowski coil is in the derivative regime. This can be useful for triggering. As such, the company PicoLAS applies a Rogowski coil in their high current fast pulser models.
The operational amplifier THS 3091 is chosen for its wide bandwidth (210~MHz), high slew rate (7300 V/μs) and wide power supply range (10 - 30~V). For convenience a single supply configuration is chosen.
Considering that the miniature coil can be positioned anywhere, the absolute scaling of the Rogowski output has to be calibrated by using the sense resistor.
A Rogowski coil is galvanically isolated from the pulser and thus eliminates the problem of “ringing”, which is caused by the very high currents in combination with a finite impedance of the ground plane. It should be noted that using a differential high frequency probe directly across R\textunderscript{sense} also can solve this problem.

The TTL input (lower part of the schematic in figure \ref{fig:driver}) is first buffered to reshape the input pulse to a fast ($<$10~ns rise time) 5~V pulse, which can be used as input for the next stage and as a buffered output.
We chose to implement a nanosecond pulse width generator, which consists of only two integrated circuits (\citet{Williams2004}).
The EL7104 MOSFET driver has been chosen for its very high speed and completes the drive circuitry. 
For future developments LEDs could be built with integrated drivers consisting of a MOSFET, capacitor bank and current sense circuitry. An example of such a design is the low cost pulsed laser diode SPL LL90\_3.
The behaviour of the LED as function of charge voltage for the SP-05-L1 LED, and a 129~ns optical pulse is shown in figure \ref{fig:LED_voltage}. For the present study it is necessary to quantify the optical output of the LED at a certain distance. The behavior of the LED is independent of the distance apart from a scaling factor. Therefore the light output is reported in scaled Watts. The light output has been measured with a BPX65 photodetector. 
\begin{figure}[!htbp]
\centering
  \includegraphics[width=0.5\textwidth]{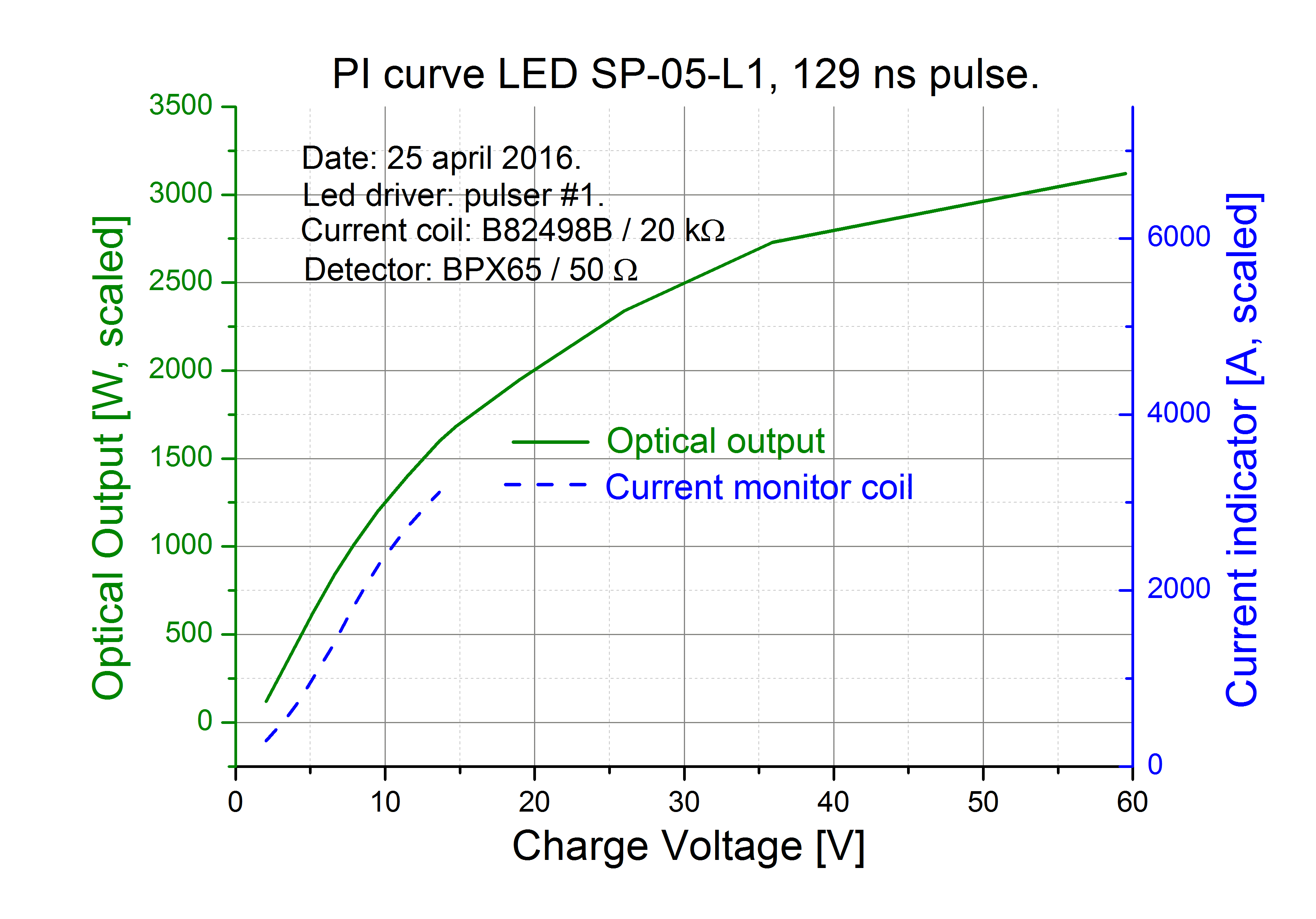}
\caption{Optical power as a function of the charge voltage of the deployed LED, 129~ns optical pulse.}
\label{fig:LED_voltage}       
  \includegraphics[width=0.5\textwidth]{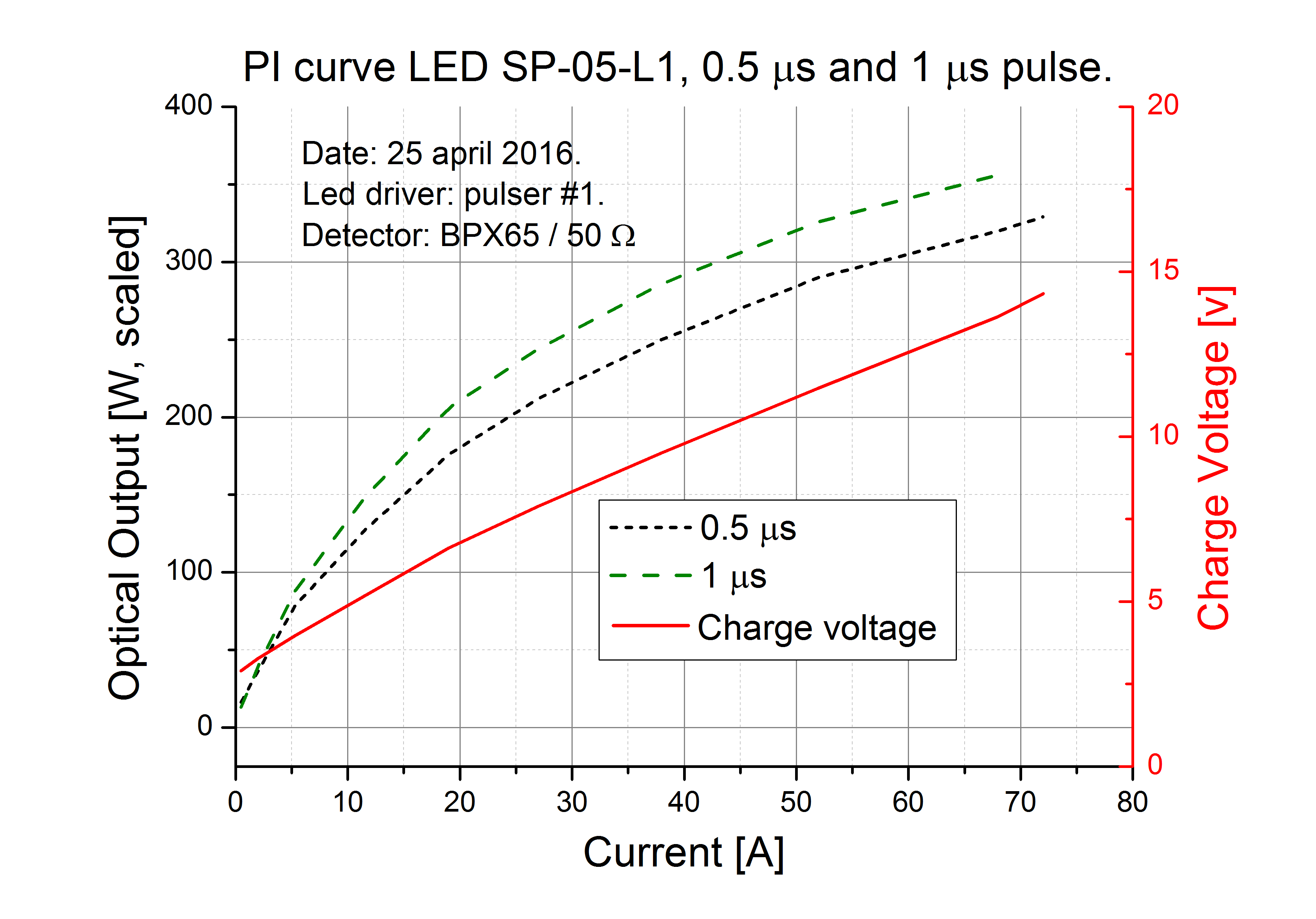}
\caption{Power and charge voltage as a function of the current for the deployed LED, 0.5~$\mu$s and 1~$\mu$s pulse.}
\label{fig:LED_current}       
\end{figure}

The current through the LED can be measured in a number of ways. Most straightforward is monitoring the voltage over R\textunderscript{sense} on an oscilloscope, as is applied to obtain figure \ref{fig:LED_current}. However, the combination of very high currents and very short pulse durations results in excessive ringing over R\textunderscript{sense}. 
Therefore, instead of the current, the charge voltage, which relates linearly to the drive current (see figure \ref{fig:LED_current}), has been chosen to estimate the optical power at high drive currents.
To estimate the current in case of the 129~ns pulse width, the 0.5~$\mu$s pulse was shortened at a charge voltage of 5~V, well in the linear regime, to 129~ns. The peak optical output dropped by a factor of 4.5, which can be linearly extrapolated to higher charge voltages, in our case up to 60~V (the driver limit) yielding 87~A. We did not investigate damage thresholds of the LED nor life time reduction at very high currents in combination with very short pulse durations. In the performed experiments we applied a more conventional charge voltage of 15~V, resulting in a current of 18~A and avoiding pulse broadening, as explained later. The momentary current in figure \ref{fig:optical_output} has been obtained by the discussed Rogowski coil.

For completeness another indirect and very simple way of estimating the  pulse current is to be mentioned. In case of repetitive signals, the average extra current supplied by the charge voltage supply due to the TTL input signal, divided by the duty cycle obtained from an optical detector, gives the pulse current. In the case of very short pulses, when the ``tail'' (see figure \ref{fig:optical_output}) is relevant in terms of the pulse width, the duty cycle may be compensated (shortened) for this.

\begin{figure}[!b]
\centering
 \includegraphics[width=0.5\textwidth]{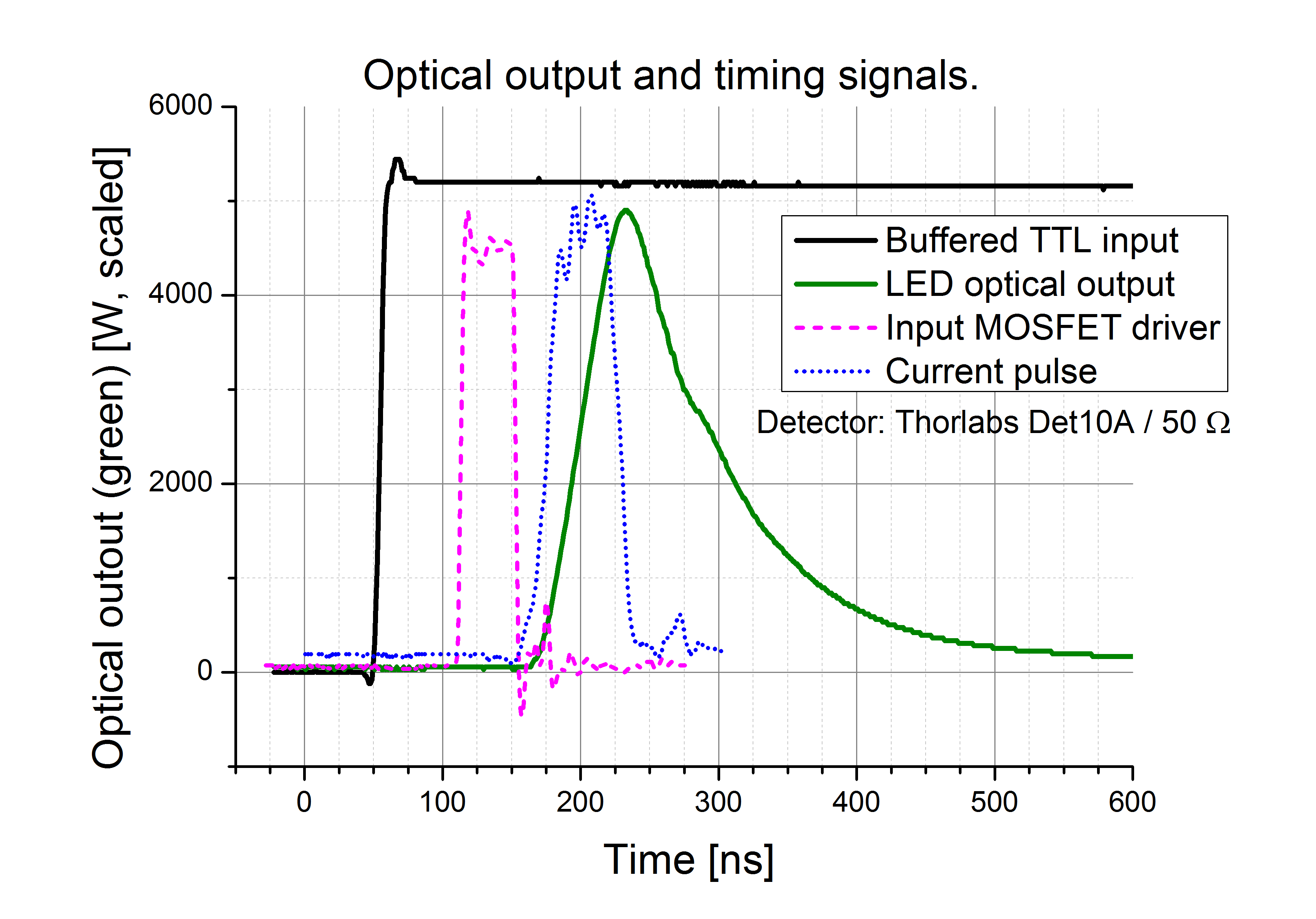}
\caption{Optical output and timing signals of the deployed LED, 129~ns optical pulse, charge voltage 5.17~V.}
\label{fig:optical_output}       
\end{figure}

From figure \ref{fig:optical_output}, a pulse width of 129~ns for the optical pulse is estimated. This value is somewhat arbitrary because a definition for pulse width has to be chosen. In this case, the effective pulse width is applied, which corresponds to the length of a pulse with a rectangular temporal profile that has the same maximum intensity and the same energy integrated over the area of the recorded curve.\\

Effective Pulse Width = Integrated temporal intensity / Maximum intensity.\\

The pulse in figure \ref{fig:optical_output} is the shortest pulse possible with this LED, considering the relatively long tail after switching off the current as well the absence of a plateau at the maximum current. In principle, it is possible to drive at higher voltages, however, the pulse width increases as well. For example, at a high charge voltage of 60 V, the pulse widens to 189~ns. The decay in time of the LED optical output intensity is related to the photophysical properties of the active material, in particular the lifetimes of electrons and holes. 

\section{Results}
\subsection{Schlieren setup}
Studying the injection region surrounding a 2~mm diameter jet in a supersonic cross flow demands a high spatial resolution. In addition the occurring velocities in the flow field are mainly supersonic (up to 480~m/s) and require a high temporal resolution for time instantaneous images. Resolving detail in the present flow field required both higher spatial and temporal resolution than achieved in earlier studies in literature utilizing a LED driven Schlieren system. Earlier works (\citet{Willert2010,Willert2012,Wilson2015}) reported a  spatial resolution of 200~$\mu m$ by 200~$\mu m$ per pixel in combination with a minimum illumination pulse of 0.5~$\mu s$. They  aimed at studying the feasibility of using LEDs as Schlieren illumination. We have pushed the concept from feasibility to accuracy for high-speed flow by increasing both the spatial as well as the temporal resolution simultaneously, while full frame images are captured.
 
The sCMOS camera together with a 100~mm Zeiss lens captured two consecutive Schlieren images with an interframe time dt =2$~\mu$s,  at a spatial resolution of 74 $\mu$m x 74 $\mu$m per pixel for an illumination pulse of 129 ns. The free stream moves less than 1 pixel width during the light pulse. 
The SA7 camera in combination with a 100~mm lens can acquire Schlieren images with a spatial resolution of 109 $\mu$m x 109 $\mu$m per pixel for an illumination pulse of 300 ns. With these settings, the free stream moves just over 1 pixel width during the illumination pulse. The settings resulted in intenser illumination compared to the measurements made with the sCMOS camera. Therefore, the Schlieren images, captured with the SA7 camera in combination with the 100 mm lens, were captured with better quality and revealed more detail of smaller density gradients, such as smaller turbulent flow structures. The SA7 in combination with a 400~mm lens resulted in a spatial resolution of 42 $\mu$m by 42 $\mu$m per pixel. However, the light pulse had to be increased to 1$~\mu$s in order to obtain sufficient illumination. In that case the free stream moved 11 pixels during the illumination, resulting in some motion blur of free stream structures. However, as the barrel shock structure changes occur at a larger time scale they were still captured sharply. These settings resolved a highly detailed view of the shock shapes for jets with a relatively low stagnation pressure and thus small flow structures. All experiments were conducted with a cutoff rate at the knife edge of 85\% $\pm$5\%.

\begin{figure}
\centering
\includegraphics[width=0.5\textwidth]{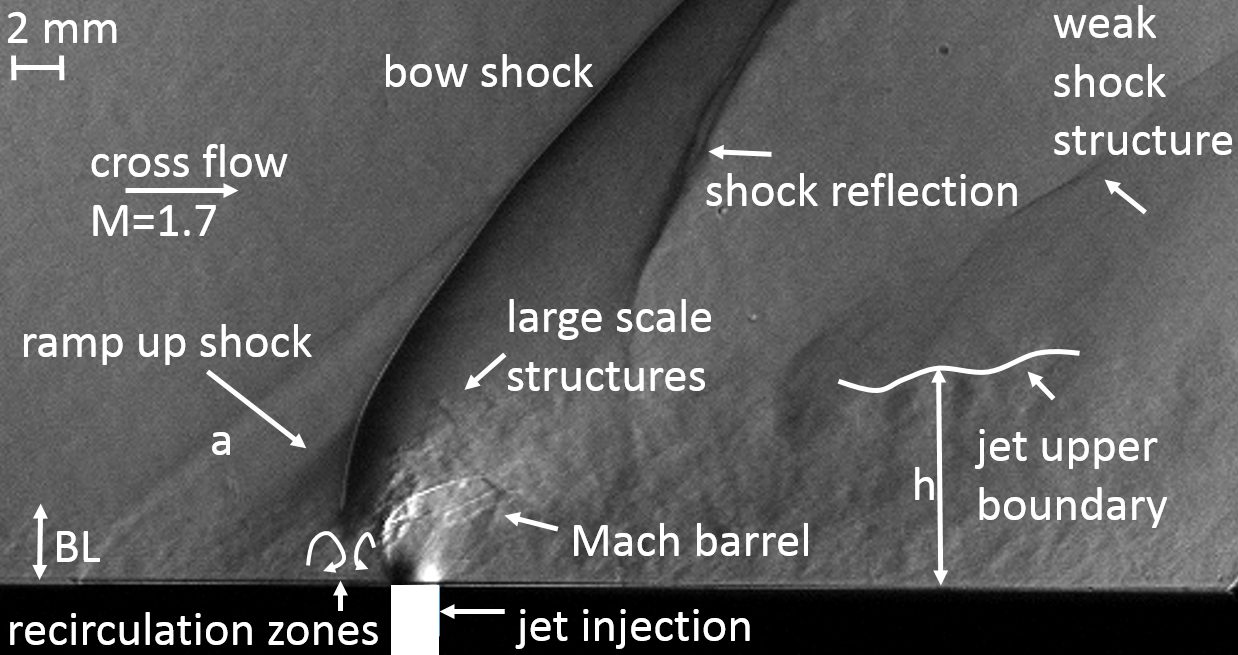}
\caption{Time-instantaneous distributions of the field of the horizontal component of the density gradient of the interaction of an underexpanded sonic jet, injected into the M=1.7 supersonic cross flow for $J=3.1$. The flow direction is from left to right. The flow structures are outlined. The sonic jet expands through a tilted Mach barrel into the cross flow, a jet core with mixing layer emerges containing large scale structures at the windward side of the jet, in which h is the jet penetration height. The bow shock emerges due to the flow obstruction by the jet, also the upstream boundary layer grows in thickness by the adverse pressure gradient in the subsonic part of the boundary layer (BL) inducing a ramp up shock. Two recirculation regions exist upstream of the jet. Apart from well known flow structures, in the present research also i) a weak shock structure is visible ii) the 3D bow shock reflection at the glass side wall and iii) Mach line at position (a) is generated at the wall in front of the jet/shock structure, which does not affect the flow field significantly. (Photron SA7 CCD, lens $100$ $mm$, illumination pulse $300$ $ns$, resolution $109$ $\mu m$ $\times$ $109$~$\mu m$ per pixel).}
\label{fig:flow_struc}       
\end{figure}

\subsection{Jet in supersonic cross flow}
An underexpanded sonic jet emanates from a 2 mm diameter orifice into the M=1.7 cross flow at reservoir pressures of 100~kPa, 200~kPa and 400~kPa. The reservoir pressures have been varied to study changes in the flow field as function of the pressure in the jet reservoir. The reservoir pressure was measured during experiments. Both the field of the horizontal component of the density gradient and that of the vertical component of the density gradient are visualized. 
A similarity parameter for an underexpanded jet injected into a supersonic cross flow is the jet-to-cross flow momentum ratio $J$, which assumes perfect gas:
\begin{equation}
J= \frac{\gamma_j p_j \mathrm{M}_j^2}{\gamma_{cf} p_{cf} \mathrm{M}^2_{cf}}
\end{equation}
in which M is the Mach number, $p$ is the static pressure and $\gamma$ is the ratio of the specific heats, which is 1.4 for air for the present range of wind tunnel temperatures. The subscripts $j$ and $cf$ indicate the jet and cross flow quantities, respectively. The static jet pressure is defined as the jet pressure at the orifice. The occurring flow structures, that are widely agreed upon (\citet{Mahesh2013,Ben-Yakar2006,Gamba2014}), of the interaction field are outlined in figure \ref{fig:flow_struc} in a captured Schlieren image resolving the field of horizontal density gradients for a $J=3.1$ jet.
\begin{figure*}[!htbp]
\centering
 \includegraphics[width=1\textwidth]{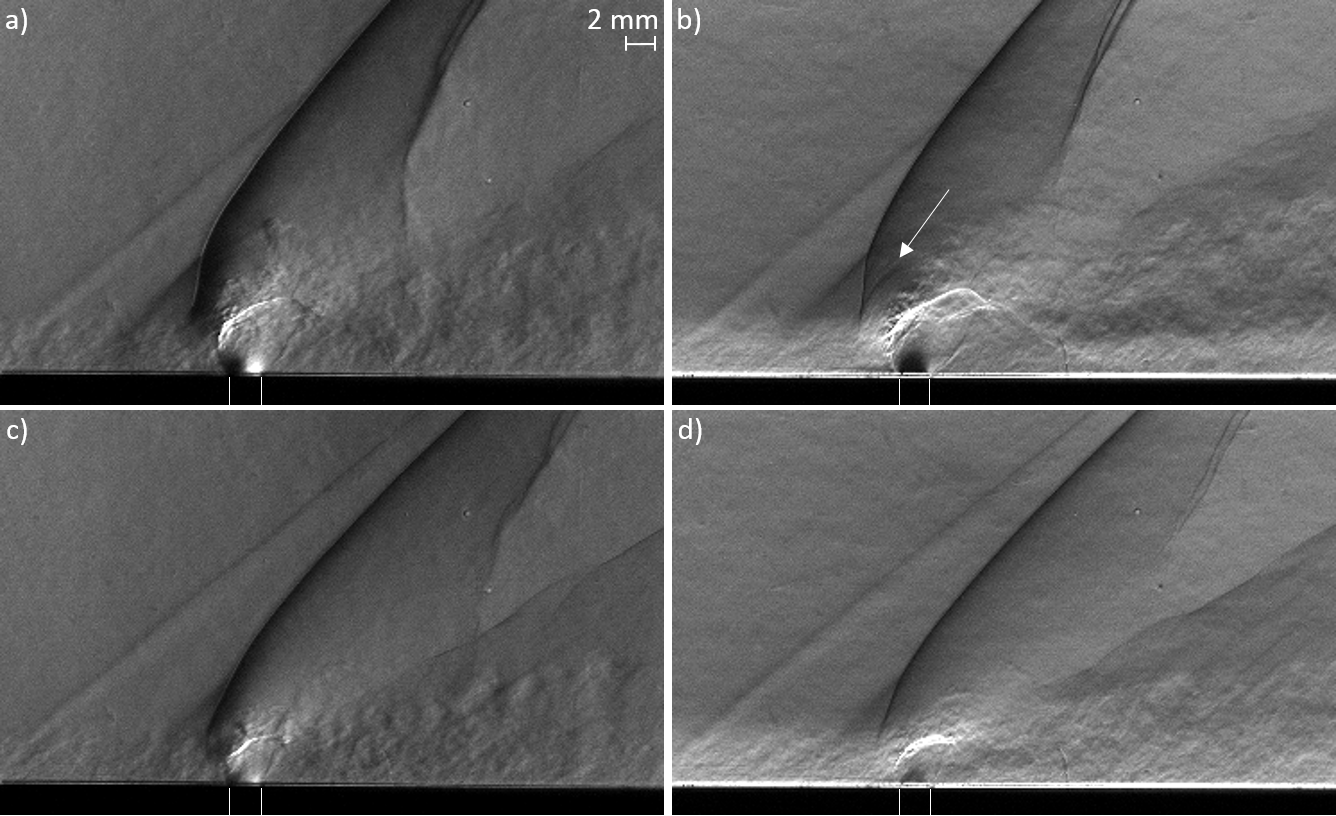}
\caption{Time-instantaneous distributions of the field of horizontal (left) and vertical (right) component of the density gradients of the interaction of an underexpanded sonic jet, injected into the $M$=1.7 supersonic cross flow. a) and b) $J=3.1$, c) and d)  $J=1.4$. The injection orifice is outlined by white lines. (Photron SA7 CCD, lens $100$ $mm$, illumination pulse $300$ $ns$, resolution $109$ $\mu m$ $\times$ $109$~$\mu m$ per pixel).}
\label{fig:instant}       
\end{figure*}

Horizontal and vertical components of the density gradient for a $J=3.1$, and $J=1.4$ jet are shown in figures \ref{fig:instant}a-d. The overall quality and resolution of the captured flow is high, while motion blur is virtually absent, clearly demonstrating the capability of the developed setup. All shock structures are visible, including a sharp image of the Mach barrel. Also, turbulent flow structures close to the Mach barrel are well resolved. Further downstream, as the fluid is mixed, the turbulent flow structures become less visible. For larger values of $J$ the turbulent structures are better visible because of higher penetration of the jet out of the boundary layer. Some images for $J=3.1$  show pockets of fluid high in the cross flow as is visible in figure \ref{fig:instant}a. In the images revealing the vertical component of the density gradient also the boundary layer is clearly visible, see figures \ref{fig:instant}b,d.

\subsubsection{Large-gradient wave}
Besides showing known flow features with a higher resolution, we also observe a new phenomenon. A high gradient wave of circular shape appears in some of the images of the distribution of the vertical component of the density gradient, in figure \ref{fig:instant}b indicated by the arrow. It appears as a disturbance originating from a point on the Mach barrel. Between the Mach barrel and the bow shock a region with subsonic flow exists where information can travel upstream. Whenever this wave interacts with the bow shock a large variation in the shape of the bow shock  is observed. The wave is also captured in the time-correlated Schlieren images. Hitherto this wave was not observed in experimental studies, and it is thanks to our temporal and spatial resolution and Schlieren sensitivity that it is now revealed.
It was only seen before in some numerical studies (\citet{Genin2010,Rana2011}) for which we thus provide validation. These numerical studies are based on (\citet{Santiago1997,Everett1998,Vanlerberghe2000}). Compared to the experiments in the present publication, the numerical studies were for a similar free stream Mach number, stagnation temperature, and $J$, though twice as large pressures and jet diameter. Besides this large-gradient wave, the  agreement between the numerical results and  our experiments is also quite good. Theory presented in these numerical studies suggests that fluctuations in the recirculation zones just upstream of the jet (figure \ref{fig:flow_struc}) are the origin of the distortions in the shape of the Mach barrel which induce such high gradient waves as the one captured in the present study. Yet, a combined experimental and numerical study should be conducted to further investigate the details of this phenomenon. We suggest that the large-gradient wave might indicate the existence of  jet screech. Screech is a feedback loop between the triple point of the Mach barrel and the lip of the jet orifice (\citet{Raman1999}). Screech as seen in sonic underexpanded jets in still air could exist in the area with subsonic flow between the wall, the Mach barrel and the normal part of the bow shock. Screech can explain the observed distortions of the Mach barrel discussed in the next paragraph. 
\begin{figure*}[!htbp]
\centering
 \includegraphics[width=1\textwidth]{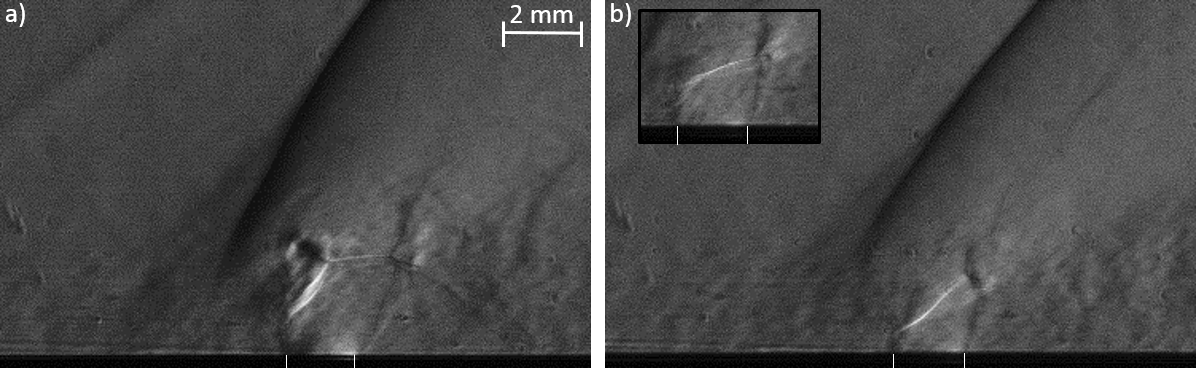}
\caption{Field of the horizontal component of the density gradient, zoomed in on the Mach barrel.  a) $J=1.4$, and b) $J=0.8$. The insert b) represents the image at 
 a different time instance to indicate the strong Mach barrel deformations in time. The injection orifice is outlined by white lines. (Photron SA7, lens $400$ $mm$, illumination pulse $1$ $\mu s$, resolution $42$ $\mu m$ by $42$ $\mu m$ per pixel)}
\label{fig:instant_zoom}       

 \includegraphics[width=1\textwidth]{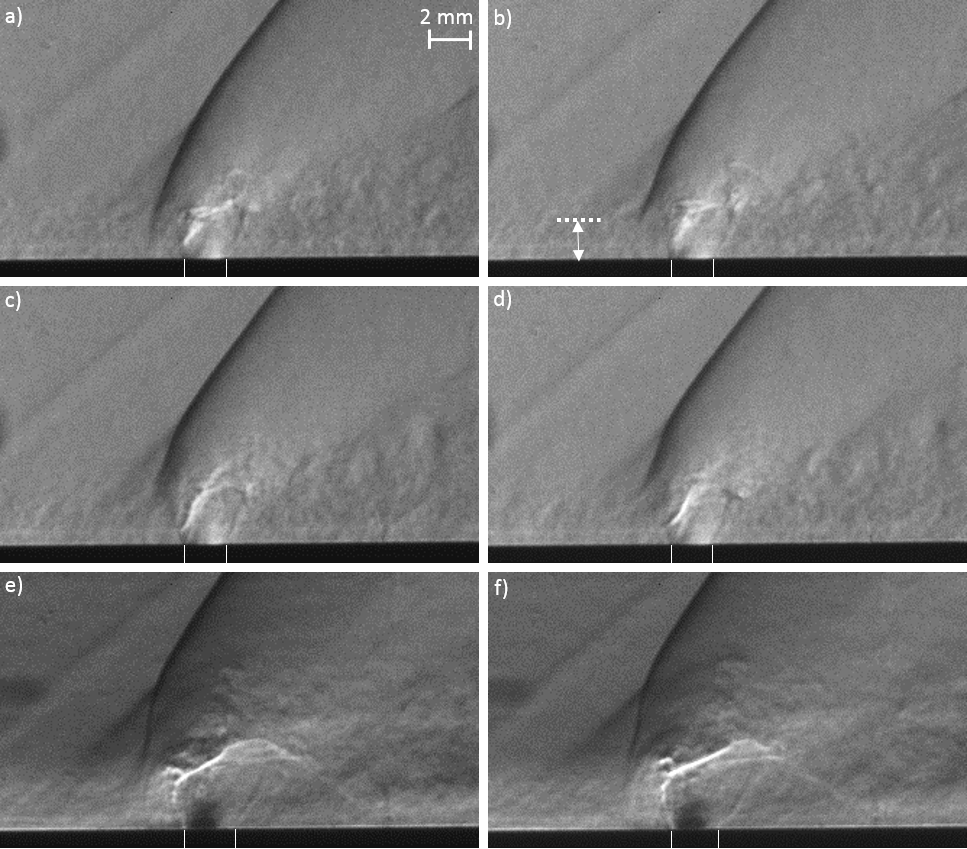} 
\caption{Time-correlated Schlieren images of a sonic jet in supersonic cross flow. The images of the left and right columns have an interframe time of 2 $\mu$s. $J=1.4$ for a) b) c) and d) showing the field of the horizontal component of the density gradient and $J=2.6$ for e) and f) showing the field of the vertical component of the density gradient. Image b) indicates the height at which the bow shock starts in the recirculation zone just upstream of the jet, which is a measure of the size of the recirculation zone. The height varies largely even between image a) and b) indicating rapid changes in the size of the recirculation zone, even within a time scale of microseconds. The injection orifice is outlined by white lines. (pco pco.edge 5.5 sCMOS, lens $100$ $mm$, illumination pulse $129$ $ns$, resolution $74$ $\mu m$ $\times$ $74$ $\mu m$ per pixel).}
\label{fig:time_correlated}       
\end{figure*}

\subsubsection{Mach barrel shock}
\label{sec:machbarrelshock}
The structure of the Mach barrel shock is clearly visible in all images for all values of $J$, even for the smaller values (see figures \ref{fig:instant},\ref{fig:instant_zoom}). Earlier work using Schlieren imaging did not always visualize a Mach barrel (\citet{Papamoschou1993,Ben-Yakar2006}). It demonstrates the advantages of using LEDs in visualizing key flow features and the resolution obtainable with the developed setup. The size of the Mach barrel increases with increasing jet pressure, because the underexpansion ratio is increased (\citet{Ashkenas1966,Orescanin2010}). The Mach barrel is skewed and ends in a tilted normal shock for larger values of $J$ (see figure \ref{fig:instant}a,b for $J=3.1$). For the lower values of $J$, $J=1.4$ (figure \ref{fig:instant}c,d and figure \ref{fig:instant_zoom}a) and $J=0.8$ (figure \ref{fig:instant_zoom}b), this normal shock is not clearly distinguished.

The shape of the Mach barrel varies in time. For smaller values of $J$ the shape is increasingly distorted (deformed) as was also observed earlier (\citet{Gamba2014}). The time correlated images reveal the time scales at which these deformations occur, see figure \ref{fig:time_correlated}, and in particular emphasized by the insert of an image at a different time instance in figure \ref{fig:instant_zoom}b. The deformation of the Mach barrel is most visible at the shock at the windward side. The deformation of this shock involves an oscillating position, sharp corners and varying shock angles.

Also the turbulent flow structures (large scale structures) surrounding the Mach barrel shock are captured, best visible in figure \ref{fig:instant}a,b. They are more distinct for higher values of $J$, because of the higher penetration of the jet out of the boundary layer. For $J$~=~3.1 in figure \ref{fig:instant}b the images show a corrugated pattern. Once in a while, a pocket of fluid is pushed high into the cross flow, while most jet fluid bends fast downstream after passing through the Mach barrel, see e.g. figure \ref{fig:instant}a. The large scale structures are important for the mixing process and thus for example for efficient fuel injection in scramjet engines. To date, the origin of these large scale structures is not yet fully known. A distortion in the windward side of the Mach barrel shock demonstrates the presence of a large fluctuation in the flow field, which could be the key in understanding these large scale structures.

\subsubsection{Bow shock}
\label{sec:bowshock}
The bow shock steepens for larger values of $J$, because the jet imposes a larger obstruction to the flow, eventually resulting in a partly normal shock (figures \ref{fig:instant}a,b and figures \ref{fig:time_correlated}e,f). Also, the bow shock is observed to be unsteady in shape and position for a fixed value of J. Unsteadiness in the shock angle is expected, because the jet does not impose a steady obstruction to the incoming supersonic cross flow. The shock angle is a function of the downstream obstruction or pressure and since the jet fluctuates so does the shock angle. Furthermore, the bow shock starts in the recirculation zone at the height at which the incoming flow velocity is supersonic. For the same experiment this height strongly varies in time, see figure \ref{fig:time_correlated}a,b. The height indicated by a dashed line in \ref{fig:time_correlated}b is much larger than the height of the end of the bow shock in figure \ref{fig:time_correlated}a.  As the images have an interframe time of only 2 $\mu$s this indicates large variations and very rapid changes in size of the recirculation zone upstream of the jet. 


Finally, a weak shock structure is observed which steepens for increasing values of $J$, as indicated in figure \ref{fig:flow_struc}. This shock structure could be a recompression shock at the lee-ward side of the jet due to re-attachement of the boundary layer as described in the experimental works of \citet{Ben-Yakar2006} and \citet{Gamba2014}.
\begin{figure}
\includegraphics[width=0.50\textwidth]{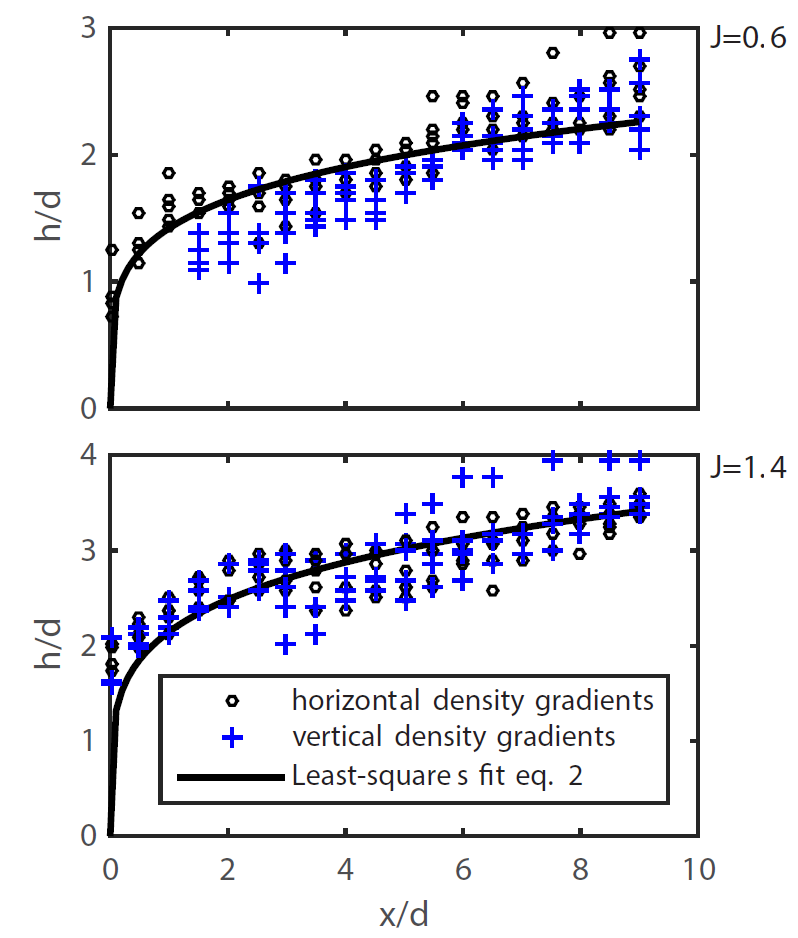} 
\caption{Jet penetration heights obtained in the present study from 10 Schlieren images (5 vertical gradient images, blue +, and 5 horizontal gradient images, black o) for the cases of a 2~mm diameter sonic jet injected into a supersonic cross flow of M=1.7 for $J$~=~0.6 and $J$~=~1.4. The measurement uncertainty corresponds to the size of the used symbols. The constants $a$, $b$ and $c$ in equation \eqref{eq:fit} are determined using a least-squares fit of this equation to the data shown above. The result of this fit is shown as a solid black line. (The position $x/d$~=~0.5 corresponds with the jet orifice center line}
\label{fig:penetration_relation}       
\end{figure}

\subsection{Jet penetration}
An important indicator for good mixing of the jet fluid with that of the free stream is the jet penetration height $h$, defined as shown in figure \ref{fig:flow_struc}. The time-instantaneous images obtained with the SA7 camera with a 100 mm zoom lens are used for determining the jet penetration height. As was also shown earlier (\citet{Gamba2014,Ben-Yakar2006}), we see that i) most of the jet fluid bends in free stream direction at the height of the Mach barrel, and ii) that further downstream the jet grows at a steady rate. At every half jet diameter ($x/d$~=~0.5) the jet penetration height $h$ was determined by visual means, using images of the fields of the horizontal and the vertical component of the density gradient. The results are shown in figure \ref{fig:penetration_relation}. We detected a similar height $h$ from the fields of the horizontal component and vertical component of the density gradient, demonstrating that detecting the upper height of these 3-dimensional structures can be done with either density gradients. The jet penetration height increases at larger distance from the jet injection location. We observe that the maximum jet height in the far field increases weakly with increasing $J$.
Many correlations for the jet penetration have been proposed to date  (\citet{Portz2006,Gruber1997,Rothstein1992,Billig1966}). Most correlations solely depend on $J$ and are expressed in the form
\begin{equation} \label{eq:fit}
\frac{h}{d}=a J^b \left( \frac{x}{d} \right) ^c,
\end{equation}
in which $h$ is the vertical jet penetration as defined in figure \ref{fig:flow_struc}, $x$ is the distance along the wall measured from the upstream (left) side of the jet orifice and $d$ is the diameter of the orifice. The dimensionless parameters $a$, $b$ and $c$ in equation (\ref{eq:fit}) are fitting parameters, which differ significantly for various studies, see e.g.\ \citet{Mahesh2013} for an overview. In other words, the flow field not only depends on $J$, but apparently also on other parameters, which are taken into account through the fitting parameters. Therefore no single correlation reconciles all experimental work (\citet{Portz2006}). For our results, the best least-squares fit for the given correlation is obtained with $a=1.82$, $b=0.49$ and $c=0.21$; these values are similar to values found before (\citet{Rothstein1992,Gruber1997,Vranos1964,Rogers1971}).

\section{Conclusion}

The resolution obtained in our experiments shows that the developed setup produces high quality Schlieren images in which the relevant flow features are captured. The LED driver is easy-to-adjust for use and optimization in several types of setups. It shows a robust performance. Illumination pulses are short (down to 130 ns). In the present supersonic flow field interaction relevant flow structures can be captured sharply. Compared to earlier work, the spatial and temporal resolution is improved significantly. 

The Schlieren images captured all known flow features present in the jet injection. The images have a level of resolution adequate for use in validation of numerical or analytical results. The shape of the Mach barrel shock was revealed for all studied jets, and the time correlated images clearly showed the time variations of the shape of the Mach barrel. In addition to much improved observation of the known flow phenomena, the experiments revealed the existence of a high gradient wave traveling upstream in the windward subsonic flow region between the Mach barrel and the bowshock. This wave was never before observed in experiments. Its existence, previously indicated by numerical studies, has now been proven. Finally, the obtained result for the jet penetration dept could be well described by the power law relation most used in literature.

The results of the present study show that the designed Schlieren setup using short pulsed LEDs with high current is very well suited to study high speed flows. The visualisation technique is powerful and we expect this study to stimulate more experimental research on other high speed (compressible) flows.

Future work includes; further developing the Schlieren setup to (further) increase the sensitivity of the setup by a quantitative research on pulsing LEDs with high-power and short duration; a study on the existence of jet screech and therewith the cause of the unsteady behaviour of the jet; and a study on parameters determining the penetration height.

\begin{acknowledgements}
We thank Harry Hoeijmakers, Mico Hirschberg, Chao Sun and Annemarie Huijser for various stimulating discussions. We thank Herman Stobbe and Steven Wanrooij for their continuous technical support. The financial support for the wind tunnel by the TKH group, represented by Alexander van der Lof, is gratefully acknowledged.
\end{acknowledgements}



\begin{thebibliography}{28}
\providecommand{\natexlab}[1]{#1}
\providecommand{\url}[1]{\texttt{#1}}
\expandafter\ifx\csname urlstyle\endcsname\relax
  \providecommand{\doi}[1]{doi: #1}\else
  \providecommand{\doi}{doi: \begingroup \urlstyle{rm}\Url}\fi

\bibitem[Argueso et~al.(2005)Argueso, Robles, and Sanz]{Argueso2005}
Marta Argueso, Guillermo Robles, and Javier Sanz.
\newblock {Implementation of a Rogowski coil for the measurement of partial
  discharges}.
\newblock \emph{Rev. Sci. Instrum.}, 76\penalty0 (6):\penalty0 6--13, 2005.

\bibitem[Ashkenas(1966)]{Ashkenas1966}
Harry Ashkenas.
\newblock {The structure and utilization of supersonic free jets in low density
  wind tunnels}.
\newblock In \emph{Proc. 4th Int. Symp. Rarefied Gas Dyn.}, pages 84--105, Vol.
  2, 1966.

\bibitem[Ben-Yakar et~al.(2006)Ben-Yakar, Mungal, and Hanson]{Ben-Yakar2006}
Adela~Ben-Yakar, M.G. Mungal, and R.K. Hanson.
\newblock {Time evolution and mixing characteristics of hydrogen and ethylene
  transverse jets in supersonic crossflows}.
\newblock \emph{Phys. Fluids}, 18:\penalty0 026101, 2006.

\bibitem[Bertin and Cummings(2003)]{Bertin2003}
John~J. Bertin and Russell~M. Cummings.
\newblock {Fifty years of hypersonics: Where we've been, where we're going}.
\newblock \emph{Prog. Aerosp. Sci.}, 39:\penalty0 511--536, 2003.

\bibitem[Billig and Schetz(1966)]{Billig1966}
F.S. Billig and J.A. Schetz.
\newblock {Penetration of gaseous jets injected into a supersonic stream.}
\newblock \emph{J. Spacecr. Rockets}, 3\penalty0 (11):\penalty0 1658--1665,
  1966.

\bibitem[Everett(1998)]{Everett1998}
D. E. Everett, M.A. Woodmansee, J.C. Dutton, and M.J. Morris.
\newblock{Wall pressure measurements for a sonic jet injected transversely into a supersonic crossflow}.
\newblock \emph{ J. Propul. Power}, 14(6) \penalty0 861–868,1998

\bibitem[Fulton et~al.(2014)Fulton, Edwards, Hassan, McDaniel, Goyne, {Rockwell
  Jr.}, Cutler, Johansen, and Danehy]{Fulton2014}
Jesse~A. Fulton, Jack~R. Edwards, Hassan~A. Hassan, James~C. McDaniel,
  Christopher~P. Goyne, Robert~D. {Rockwell Jr.}, Andrew~D. Cutler, Craig~T.
  Johansen, and Paul~M. Danehy.
\newblock {Large-Eddy/-Averaged Navier–Stokes Simulations of Reactive
  Flow in Dual-Mode Scramjet Combustor}.
\newblock \emph{J. Propuls. Power}, 30\penalty0 (3):\penalty0 558--575, 2014.

\bibitem[Gamba et~al.(2014)Gamba, Miller, and Mungal]{Gamba2014}
Mirko.~Gamba, Victor~A. Miller, and M.Godfrey Mungal.
\newblock {The reacting transverse jet in supersonic crossflow: physics and
  properties}.
\newblock In \emph{AIAA 19th Int. Sp. Planes Hypersonic Syst. Technol. Conf.},
  Atlanta, 2014. AIAA Aviation.

\bibitem[G\'{e}nin and Menon(2010)]{Genin2010}
Franklin G\'{e}nin and Suresh Menon.
\newblock {Dynamics of sonic jet injection into supersonic crossflow}.
\newblock \emph{J. Turbul.}, 11\penalty0 (4):\penalty0 1--30, 2010.

\bibitem[Gruber et~al.(1997)Gruber, Nejad, Chen, and Dutton]{Gruber1997}
Mark~R. Gruber, A.S.S. Nejad, T.H.H. Chen, and J.~Craig Dutton.
\newblock {Compressibility effects in supersonic transverse injection
  flowfields}.
\newblock \emph{Phys. Fluids}, 9:\penalty0 1448--1461, 1997.

\bibitem[Mahesh(2013)]{Mahesh2013}
Krishnan Mahesh.
\newblock {The Interaction of Jets with Crossflow}.
\newblock \emph{Annu. Rev. Fluid Mech.}, 45:\penalty0 379--407, 2013.

\bibitem[Morimoto et~al.(2016)Morimoto, Yamashita, Tabata, Aso, and
  Tani]{Morimoto2016}
Naoki Morimoto, Jun Yamashita, Akihiko Tabata, Shigeru Aso, and Yasuhiro Tani.
\newblock {Schlieren Visualization Technique for High-Enthalpy and Low-Density
  Flow with LED Light Source}.
\newblock In \emph{54th AIAA Aerosp. Sci. Meet.}, pages 1--11, San Diego, 2016.
  AIAA Sci Tech.

\bibitem[Orescanin and Austin(2010)]{Orescanin2010}
Mara~M. Orescanin and J.~M. Austin.
\newblock {Exhaust of Underexpanded Jets from Finite Reservoirs}.
\newblock \emph{J. Propuls. Power}, 26\penalty0 (4):\penalty0 744--753, 2010.

\bibitem[Papamoschou and Hubbard(1993)]{Papamoschou1993}
Dimitri~Papamoschou and D.~G. Hubbard.
\newblock {Visual observations of supersonic transverse jets}.
\newblock \emph{Exp. Fluids}, 14\penalty0 (6):\penalty0 468--476, 1993.

\bibitem[Parziale(2015)]{Parziale2015}
Nick J. Parziale, Bryan E. Schmidt, P. Wang, H. Hornung, J. Shepherd.
\newblock{Pulsed Laser Diode for Use as a Light Source for Short Exposure, High-Frame-Rate Flow Visuaization}. 
\newblock \emph{53rd AIAA SciTech 2015}, 4-9 January, Kissimmee, Florida. \# AIAA 2015-0530, 2015.

\bibitem[Portz and Segal(2006)]{Portz2006}
Ron Portz and Corin Segal.
\newblock {Penetration of Gaseous Jets In Supersonic Flows}.
\newblock \emph{AIAA J.}, 44\penalty0 (10):\penalty0 2426--2429, 2006.

\bibitem[Raman(1999)]{Raman1999}
Ganesh~Raman.
\newblock {Supersonic jet screech: Half-century from Powell to the present}.
\newblock \emph{J. Sound Vib.}, 225\penalty0 (3):\penalty0 543--571, 1999.

\bibitem[Rana et~al.(2011)Rana, Thornber, and Drikakis]{Rana2011}
Zeeshan A. Rana, Ben~Thornber, and Dimitris~Drikakis.
\newblock {Transverse jet injection into a supersonic turbulent cross-flow}.
\newblock \emph{Phys. Fluids}, 23:\penalty0 046103, 2011.

\bibitem[{Rockwell Jr.} et~al.(2014){Rockwell Jr.}, Goyne, Rice, Kouchi,
  McDaniel, and Edwards]{Rockwell2014}
Robert~D. {Rockwell Jr.}, Christopher~P. Goyne, Brian~E. Rice, Toshinori
  Kouchi, James~C. McDaniel, and Jack~R. Edwards.
\newblock {Collaborative Experimental and Computational Study of a Dual-Mode
  Scramjet Combustor}.
\newblock \emph{J. Propuls. Power}, 30\penalty0 (3):\penalty0 530--538, 2014.

\bibitem[Rogers(1971)]{Rogers1971}
R.~Clayton Rogers.
\newblock {A Study of the Mixing of Hydrogen Injected Normal to a Supersonic
  Airstream}.
\newblock Technical report, NASA Rep. TN-D6114, NASA Langley Res. Cent.,
  Hampton, VA, 1971.

\bibitem[Rothstein and Wantuck(1992)]{Rothstein1992}
A.D. Rothstein and P.J. Wantuck.
\newblock {A Study of the Normal Injection of Hydrogen into a Heated Supersonic
  Flow Using Planar Laser-Induced Fluorescence}.
\newblock In \emph{28th Jt. Propuls. Conf. Exhib.}, Nashvile, U.S.A., 1992.

\bibitem[Rouwenhorst(2012)]{Rouwenhorst2012}
Driek~Rouwenhorst.
\newblock \emph{{Experimental Study on Pressure Oscillations Induced by
  Supersonic Flow past a Rectangular Cavity}}.
\newblock Msc thesis, University of Twente, 2012.

\bibitem[Santiago and Dutton(1997)]{Santiago1997}
Juan~Gabriel Santiago and J.~Craig Dutton.
\newblock {Velocity Measurements of a Jet Injected into a Supersonic
  Crossflow}.
\newblock \emph{J. Propuls. Power}, 13\penalty0 (2):\penalty0 264--273, 1997.

\bibitem[Settles(2001)]{Settles2001}
Gary~S. Settles.
\newblock \emph{{Schlieren and shadowgraph Technique, visualizing Phenomena in
  Transparent Media}}.
\newblock Springer-Verlag, Berlin, 2001.

\bibitem[VanLerberghe et~al.(2000)VanLerberghe, Santiago, Dutton, and
  Lucht]{Vanlerberghe2000}
Wayne~M. VanLerberghe, Juan~Gabriel Santiago, J.~Craig Dutton, and Robert~P.
  Lucht.
\newblock {Mixing of a Sonic Transverse Jet Injected into a Supersonic Flow}.
\newblock \emph{AIAA J.}, 38\penalty0 (3):\penalty0 470--479, 2000.

\bibitem[Vranos and Nolan(1964)]{Vranos1964}
Alexander Vranos and Jamers~J. Nolan.
\newblock {Supersonic Mixing of Helium and Air}.
\newblock Technical report, Bumblebee Rep. No. TG 63-53, Appl. Phys. Lab., John
  Hopkings Univ,, 1964.

\bibitem[Willert et~al.(2010)Willert, Stasicki, Klinner, and
  Moessner]{Willert2010}
Christian~E. Willert, B~Stasicki, J~Klinner, and S~Moessner.
\newblock {Pulsed operation of high-power light emitting diodes for imaging
  flow velocimetry}.
\newblock \emph{Meas. Sci. Technol.}, 21:\penalty0 075402, 2010.

\bibitem[Willert et~al.(2012)Willert, Mitchell, and Soria]{Willert2012}
Christian~E. Willert, Daniel~M. Mitchell, and Julio Soria.
\newblock {An assessment of high-power light-emitting diodes for high frame
  rate schlieren imaging}.
\newblock \emph{Exp. Fluids}, 53:\penalty0 413--421, 2012.

\bibitem[Williams(2004)]{Williams2004}
Jim Williams.
\newblock {Signal Sources, Conditioners and Power Circuitry, Linear Technology Application Note AN98}, 2004.

\bibitem[Wilson et~al.(2015)Wilson, Gustafson, Lincoln, Murari, and
  Johansen]{Wilson2015}
S.~Wilson, G.~Gustafson, D.~Lincoln, K.~Murari, and Craig~T. Johansen.
\newblock {Performance evaluation of an overdriven LED for high-speed schlieren
  imaging}.
\newblock \emph{J. Vis.}, 18:\penalty0 35--45, 2015.

\end{thebibliography}
\end{document}